%% file: twc.tex
\ifdefined\pdfminorversion\pdfminorversion=7\fi
\documentclass[10pt,letterpaper,journal,comsoc]{IEEEtran}
\usepackage{amsmath,amsfonts,amssymb,amsthm}
\usepackage{algorithmic}
\usepackage{array}
\usepackage{booktabs}
\usepackage[caption=false,font=normalsize,labelfont=sf,textfont=sf]{subfig}
\usepackage{textcomp}
\usepackage{stfloats}
\usepackage{url}
\usepackage{verbatim}
\usepackage{graphicx}
\usepackage{cite}

\newif\ifReviewLinks
\ReviewLinkstrue
\usepackage{hyperref}
\hypersetup{pdftitle={Reciprocal Range-Doppler Estimation with Partial Path Association for Asynchronous Bistatic OFDM-ISAC},pdfauthor={Jiayao Shao, Zhendong Li, Wenjie Wang, Zhou Su, Kaitao Meng, and Wen Chen}}
\ifReviewLinks
  \hypersetup{colorlinks=false,pdfborder={0 0 0.6},pdfborderstyle={},
    citebordercolor={0 1 0},linkbordercolor={1 0 0},urlbordercolor={0.35 0.65 1},
    filebordercolor={1 0 0},runbordercolor={1 0 0}}
\else
  \hypersetup{hidelinks}
\fi

\DeclareRobustCommand{\paperemail}[1]{%
  \begingroup
  \hypersetup{pdfborder={0 0 0},pdfborderstyle={}}%
  \urlstyle{same}%
  \mbox{\href{mailto:#1}{\nolinkurl{#1}}}%
  \endgroup
}

\newcounter{ExpandedMatrixSavedEquation}
\providecommand{\ExpandedMatrixNumber}{0}
\makeatletter
\newcommand{\ReserveExpandedMatrixNumber}{%
  \stepcounter{equation}
  \protected@write\@auxout{}{\string\gdef\string\ExpandedMatrixNumber{\theequation}}}
\makeatother

\newtheoremstyle{twcresult}
  {\topsep}{\topsep}{\itshape}{}{\normalfont\bfseries}{}{.5em}
  {\thmname{#1}\thmnumber{ #2}\thmnote{\normalfont\space(#3)}\normalfont:}
\theoremstyle{twcresult}
\newtheorem{proposition}{Proposition}
\newtheorem{lemma}{Lemma}

\theoremstyle{definition}

\newcommand{\CN}{\mathcal{CN}}
\newcommand{\E}{\mathbb{E}}
\newcommand{\Real}{\operatorname{Re}}
\newcommand{\diag}{\operatorname{diag}}
\newcommand{\argmin}{\operatorname*{arg\,min}}
\newcommand{\argmax}{\operatorname*{arg\,max}}
\newcommand{\tr}{\operatorname{tr}}
\newcommand{\CapBasic}{Range and Doppler RMSE in S1 ($M_A=M_B=280$). Each estimator is scored on its finite outputs for the two longer paths. Han and ADD-Cal use corrected target echoes at BS A.}
\newcommand{\CapAssociation}{Proposed receiver under four visibility patterns: (a) pair precision (solid) and recall (dashed); (b) useful recovery.}
\newcommand{\CapResources}{Effects of directional resources at $-10$ dB with fixed total energy or fixed power per symbol.}
\newcommand{\CapResidual}{RMSE under clock variation at $0$ dB for (a) timing and (b) carrier frequency offsets. The horizontal axis in (b) is linear up to 1 Hz and logarithmic above it. Scoring uses the two paths with the longer ranges.}
\newcommand{\CapAblation}{S2 association rules with common coherent recovery. (a) Fraction of accepted pairs with the correct identity. (b) Fraction of planned common targets correctly paired and within both error tolerances.}

\newcommand{\FigOneWidth}{1.00\columnwidth}

\title{\fontsize{24}{28}\selectfont
Reciprocal Range-Doppler Estimation\\
with Partial Path Association for\\
Asynchronous Bistatic OFDM-ISAC}
\author{Jiayao Shao, Zhendong Li, Wenjie Wang, Zhou Su, Kaitao Meng, and Wen Chen%
\thanks{Jiayao Shao, Zhendong Li, and Wenjie Wang are with the School of Information and Communication Engineering, Xi'an Jiaotong University, Xi'an, China (e-mail: \paperemail{shaojiayao@stu.xjtu.edu.cn}; \paperemail{lizhendong@xjtu.edu.cn}; \paperemail{wjwang@mail.xjtu.edu.cn}). Zhou Su is with the School of Cyber Science and Engineering, Xi'an Jiaotong University, Xi'an, China (e-mail: \paperemail{zhousu@ieee.org}). Kaitao Meng is with the University of Manchester, Manchester, U.K. (e-mail: \paperemail{kaitao.meng@manchester.ac.uk}). Wen Chen is with the School of Integrated Circuits, Shanghai Jiao Tong University, Shanghai, China (e-mail: \paperemail{wenchen@sjtu.edu.cn}).}}

\begin{document}

\bstctlcite{IEEEBST:referenceformat}

\makeatletter
\def\@IEEENORMtitlevspace{0.25\baselineskip}
\def\@IEEEMINtitlevspace{0pt}
\makeatother
\maketitle
\begin{abstract}
In bistatic integrated sensing and communication (ISAC), clock asynchronism shifts the delay and Doppler measured from orthogonal frequency-division multiplexing (OFDM) echoes. The two stations may see different reflectors, so the receiver must identify which echoes belong to the same path before combining the two views. For partial path association, PreVote finds a common displacement in the directional delay and Doppler differences. One-to-one assignment at this displacement accepts compatible pairs and allows other paths to remain unmatched. Each station uses coherent compression to form a subcarrier vector and a slow-time vector per path. For range-Doppler estimation in reciprocal sensing, the center multiplies paired vectors without conjugation. The opposite constant clock terms cancel, and a correct pair yields twice the physical phase slopes of one path. A Cram\'er-Rao bound (CRB) for fixed identities and phase branches quantifies the shared clock uncertainty. At the higher signal-to-noise ratios tested, reciprocal fusion yields lower range and Doppler errors than both receivers using calibrated echoes. Under partial visibility, displacement-based matching recovers more targets with correct identities and prescribed accuracy than separate centering with the same fusion stage. Balancing directional energy and extending the common aperture improve Doppler accuracy, while carrier-frequency variation increases error.
\end{abstract}

\begin{IEEEkeywords}
ISAC, clock asynchronism, partial path association, range-Doppler estimation, reciprocal sensing.
\end{IEEEkeywords}

\section{Introduction}\label{sec:introduction}
\IEEEPARstart{I}{ntegrated} sensing and communication (ISAC) enables wireless networks to sense the environment while transmitting data by sharing waveforms, spectrum, and radio infrastructure~\cite{liu2022isac,zhang2022perceptive,du2023v2iisac}. Array and time-frequency measurements in orthogonal frequency-division multiplexing (OFDM) encode angle, delay, and Doppler, enabling sensing from communication signals~\cite{dai2026tutorial}. In the full-duplex design of~\cite{li2026ma}, one base station serves uplink and downlink users and receives target echoes on its own array. Multistatic sensing extends reception to geographically separated sites, so different stations can observe the same target from different directions. The authors in~\cite{yang2026japs} jointly designed transmit beams, receive filters, and uplink power allocation so that active and passive sensing could share a multistatic network with uplink and downlink communications. In~\cite{liu2024cooperative}, simulations showed how multistation fusion improved position and velocity estimates, and a cellular field trial demonstrated cooperative sensing.

Turning those complementary views into a joint estimate requires coordinating which stations observe the scene and deciding how to combine their measurements. These decisions arise at both the network and receiver levels. By combining coordinated multipoint transmission with distributed multiple-input multiple-output (MIMO) radar, the authors in~\cite{meng2025cooperative} analyzed how localization accuracy scales with the number of cooperating stations and optimized cluster sizes and transmit power. At the receiver, each station can extract angle, range, and velocity by jointly processing its own MIMO-OFDM echoes over a coherent observation interval~\cite{xiao2024joint}. The fusion center can combine local estimates or process signal measurements jointly. When joint processing calls for more than local estimates, transferring full array channel state information (CSI) places a load on the fronthaul. In~\cite{zhang2024fronthaul}, beamforming and compression of the transmitted signals and received echoes were jointly optimized under fronthaul capacity constraints. For cooperative passive sensing, a line-of-sight (LoS) reference was used to compensate clock terms before compressing path responses into range and velocity vectors for fusion~\cite{wei2025cooperative}. In that design, the direct and reflected signals share a clock offset, so the compensated vectors carry phase differences relative to the direct path. Whether stations exchange local estimates or signal samples, independent clocks complicate their combination. A timing offset (TO) enters the same subcarrier phase slope as propagation delay, and a carrier-frequency offset (CFO) enters the same slow-time phase slope as physical Doppler~\cite{wu2024async}. Separating clock effects from propagation is therefore central to fusing observations from asynchronous stations.

Existing methods distinguish propagation from clock effects using four sources of information. Correlations and ratios exploit structure across receive channels or frequencies~\cite{ni2021cacc,hu2024casr,hu2026crossfreq}. Known signals or propagation landmarks provide a clock reference~\cite{brunner2025,jiang2024cooperation,pegoraro2024jump,li2026asynchronous}. Temporal models track offsets across observation blocks~\cite{chen2024kalman,wang2026time}, while reciprocal methods compare measurements after the stations exchange transmitter and receiver roles~\cite{han2025ota,chen2026tensor}. The choice of information determines which part of the clock ambiguity can be resolved. For a single moving path, CSI ratios suppress common phase disturbances and yield a closed-form bound on bistatic Doppler precision at high signal-to-noise ratio (SNR)~\cite{hu2024casr}. Cross-frequency correlation in~\cite{hu2026crossfreq} removes a shared phase contribution and uses a dominant LoS path to resolve the remaining timing phase. An explicit reference can instead come from active echoes, as in cooperative passive sensing~\cite{jiang2024cooperation}, or from a static reflection whose phase is tracked after aligning channel magnitude profiles~\cite{pegoraro2024jump}. Across successive blocks, discrepancies between measured and geometrically predicted LoS parameters can drive recursive updates of time-varying offsets~\cite{wang2026time}. With reciprocal measurements, conjugate products pooled over scatterers isolate the clock terms~\cite{han2025ota}, and a joint tensor model uses a shared CFO factor to couple the directions~\cite{chen2026tensor}. The same link reversal also carries information about the target. At a common epoch, its physical delay and Doppler are the same in both directions. Recovering these quantities from reciprocal echoes requires identifying which responses belong to the same path.

In cooperative sensing, data association typically uses geometric or calibration information to determine which observations belong to the same target. In~\cite{yu2025association}, directional measurements from distributed access points were associated by clustering before estimating target locations. Multipath structure in an uplink can also be used to associate direction, delay, and Doppler estimates~\cite{wang2024clutter}. Bu and Zhou~\cite{bu2023joint} jointly modeled measurement assignment and sensor biases for multisensor tracking. For aerial targets, a minimum spanning tree links estimates across base stations and removes false matches before position and velocity are estimated~\cite{tang2025lowaltitude}. In user-assisted localization, ranges are first estimated in the presence of synchronization errors, then associated with targets using measurements from the base station and selected user anchors~\cite{guo2025ue}. In reciprocal sensing, agreement in the effective delay and Doppler differences provides another association cue. When the two observations contain the same paths, the formulation in~\cite{chen2026scpd} associates the responses by permuting a fixed common list and gives a joint bound for clock and target parameters. Its extended matrix pencil comparator forms nonconjugate products of steering vectors reconstructed from local slope estimates. Once correspondence is specified, clock uncertainty and observation resources govern estimation precision. Bounds for passive sensing quantify how clock asynchronism affects the precision of angle, delay, and complex gain estimates~\cite{zhao2024bounds}. At the network level, spatial diversity, frequency bands, and observation scheduling affect position and velocity bounds~\cite{pu2026synthetic}. With partial visibility, lists of equal length can contain different paths. Reciprocal estimation then requires both choosing the common subset and identifying its corresponding responses across stations.

Motivated by the above discussion, we develop a reciprocal receiver for asynchronous bistatic OFDM sensing. At the fusion center, PreVote uses agreement among directional slope differences to estimate a shared clock displacement. Local angle, power, and fit information then helps the center choose compatible pairs and leave other responses unmatched. Each station uses a central phase reference and power weighting to compress a local path response into one vector across subcarriers and another across symbols. The association determines which of these vectors the center combines. Nonconjugate multiplication adds the slopes from the two directions and cancels their opposite constant clock terms. For a correct pair, the sum is twice the physical slope of one path on each axis. A common clock also couples the measurements of different paths, even after their correspondence is known. We use this coupling to quantify the precision available from the two directions. The main contributions are as follows.
\begin{itemize}
\item We extend reciprocal fusion to path lists with unknown overlap. PreVote uses local fit quality to score candidate clock displacements before partial assignment. We give sufficient conditions on pairing and rejection costs for the correct assignment to be unique. Local coherent compression supplies two phase vectors per path. For a correct pair, their nonconjugate products cancel the opposite constant clock terms and carry twice the physical slopes, enabling range and Doppler estimation without exchanging full array observations.
\item Starting from the raw array observations, we derive the Fisher information and Cram\'er-Rao bound (CRB) with unknown directional gains, angles, and clock offsets under fixed identities and phase branches. For isolated paths with known angles, the covariance bound reveals how common paths constrain the shared clock and how its uncertainty affects paths observed in one direction. This calculation identifies which physical contrasts retain their known-clock bound and when the half-sum of independent, efficient directional estimates attains the joint bound. It also relates directional energy and observation aperture to the attainable precision in range and Doppler.
\item The experiments separate the effect of path selection from the accuracy of reciprocal fusion. Under complete visibility, we compare vector fusion with estimates from calibrated target echoes. Across four visibility patterns, we count correctly paired targets recovered within both range and Doppler tolerances. Holding the coherent recovery stage fixed then reveals how the choice of association rule affects that count. Directional energy and symbol allocation tests relate recovery to the predicted effects of information balance and common aperture. We also test how phase evolution within the observation blocks affects Doppler accuracy.
\end{itemize}

\begingroup
\clubpenalty=10000
Section~\ref{sec:model} formulates the two directional observations for partially overlapping path sets. Sections~\ref{sec:association} and~\ref{sec:recovery} develop the displacement test and the subsequent recovery of range and Doppler. Section~\ref{sec:theory} derives identifiability conditions and estimation bounds, and Section~\ref{sec:numerical} examines their implications for association, resource allocation, and clock variation. Section~\ref{sec:conclusion} summarizes the findings.
\par
\endgroup

\emph{Notation.} Italic letters denote scalars, and bold lowercase and uppercase letters denote vectors and matrices. Hats denote estimates and $j=\sqrt{-1}$ in complex phases. Real and complex matrix spaces are $\mathbb R^{a\times b}$ and $\mathbb C^{a\times b}$. $\mathbf I$ and $\mathbf1$ are the identity matrix and all-ones vector. Transpose, conjugate transpose, and conjugation use $(\cdot)^{\mathrm{T}}$, $(\cdot)^{\mathrm H}$, and $(\cdot)^*$. The Euclidean and Frobenius norms are $\|\cdot\|_2$ and $\|\cdot\|_F$, with inner product $\langle\mathbf A,\mathbf B\rangle_F=\tr(\mathbf A^{\mathrm H}\mathbf B)$. Kronecker and elementwise products use $\otimes$ and $\odot$. We use $\diag$ for diagonal-matrix construction, $\succeq$ for positive-semidefinite ordering, and $\operatorname{rank}$ for rank. Real part, imaginary part, and principal phase are $\Real$, $\operatorname{Im}$, and $\arg$. Clipping and median are $\operatorname{clip}_{[a,b]}$ and $\operatorname{med}$. Floor and ceiling are $\lfloor\cdot\rfloor$ and $\lceil\cdot\rceil$, and $\argmin$ and $\argmax$ return sets of optima. Indicators and set cardinality use $\mathbf1_{\mathcal A}$ and $|\mathcal S|$. Expectation, variance, and covariance use $\E$, $\operatorname{var}$, and $\operatorname{cov}$. The law $\CN(\cdot,\cdot)$ is proper complex Gaussian with mean and covariance arguments.

\section{Reciprocal Observation Model}\label{sec:model}
\subsection{Propagation and Asynchronous Observations}
\begin{figure}[!t]
\centering
\includegraphics[
  width=\FigOneWidth,
  trim=10pt 40pt 10pt 20pt,
  clip
]{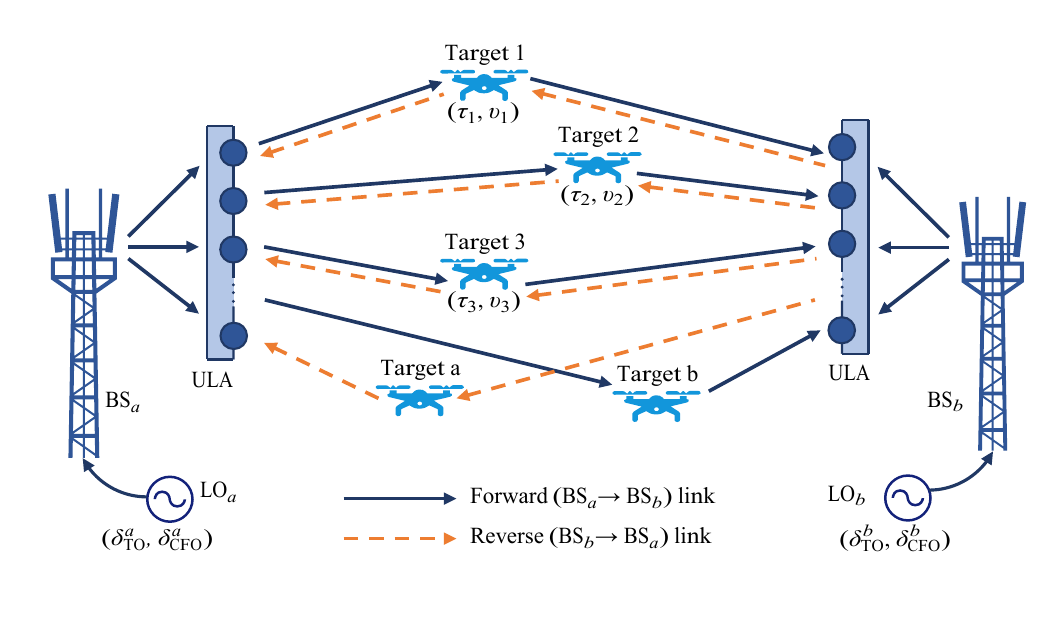}
\caption{Reciprocal bistatic sensing with common numbered paths and lettered paths visible in one direction. The local oscillators (LOs) are labeled by station.}
\label{fig:system_overview}
\end{figure}

Consider two base stations (BSs), A and B, at known locations $\mathbf b_A$ and $\mathbf b_B$, respectively. Each station has its own local oscillator (LO). The stations exchange transmitter and receiver roles in adjacent blocks. In Fig.~\ref{fig:system_overview}, $\mathrm{BS}_a$ and $\mathrm{BS}_b$ denote the stations, $\delta_{\rm TO}^{a,b}$ and $\delta_{\rm CFO}^{a,b}$ mark local offsets, and $\upsilon_\ell=\nu_\ell$. Direction $q=+1$ denotes reception at BS A and $q=-1$ reception at BS B. Each receiver uses a uniform linear array (ULA) of $N_r$ elements, $N$ active subcarriers, and $M_q$ symbols. The sets of usable paths $\mathcal L_A,\mathcal L_B$ have intersection $\mathcal L_c$ and sizes $L_A,L_B,L_c$. The receiver knows $L_A,L_B$ and infers the common subset and its identities.

For scatterer $\ell$ at $\mathbf p_\ell(t)$, the delay is $\tau_\ell=R_\ell(t_0)/c$ with reference epoch $t_0$, propagation speed $c$, and bistatic path length
\begin{equation}
R_\ell(t)=\|\mathbf p_\ell(t)-\mathbf b_A\|_2+\|\mathbf p_\ell(t)-\mathbf b_B\|_2.
\label{eq:geometry}
\end{equation}
At nominal carrier frequency $f_c$, the phase $\phi_\ell(t)=-2\pi f_cR_\ell(t)/c$ gives $\nu_\ell=\dot\phi_\ell(t_0)/(2\pi)=-f_c\dot R_\ell(t_0)/c$. Role exchange therefore preserves delay and Doppler at the common epoch. $R=c\tau$ and $\operatorname{var}(R)=c^2\operatorname{var}(\tau)$. Both blocks use the envelope delay at $t_0$ and the linearized carrier phase. Directional gains and array responses are constant within each block. Moving-node geometry enters through $\dot R_\ell$~\cite{ventura2026asymov}.

With subcarrier spacing $\Delta f$, cyclic prefix (CP) duration $T_{\rm CP}$, and transform size $K_F\geq N$, set $T_u=1/\Delta f$, $T_s=T_u+T_{\rm CP}$, and $T_{\rm samp}=T_u/K_F$. For even $N$, the active frequencies are $f_n=(n-N/2)\Delta f$. For unit-modulus pilots $p_q[n,m]$, the useful OFDM symbol is
\begin{equation}
x_{q,m}(t)=\frac{1}{\sqrt{K_F}}\sum_{n=0}^{N-1}p_q[n,m]e^{j2\pi f_nt},\quad 0\leq t<T_u.
\label{eq:ofdmwaveform}
\end{equation}
The CP supplies the periodic extension for delayed samples within the timing window. Let $\tau_o$ denote the transmit timing origin at BS B relative to the receive origin at BS A, and let $f_o=f_B-f_A$ for station carrier frequencies $f_A$ and $f_B$. Exchanging the node roles gives
\begin{equation}\label{eq:effective}
\theta_{\tau,q,\ell}=\tau_\ell+q\tau_o,\qquad
\theta_{\nu,q,\ell}=\nu_\ell+qf_o.
\end{equation}
After CP removal, the narrowband model with negligible intercarrier interference (ICI) gives the $r$th antenna samples as
\begin{equation}
\begin{aligned}
z_q[r,k,m]&\simeq\sum_{\ell\in\mathcal L_q}\alpha_{q,\ell}e^{j\pi ru_{q,\ell}}
e^{j2\pi t_{q,m}\theta_{\nu,q,\ell}}\\
&\qquad\times x_{q,m}(kT_{\rm samp}-\theta_{\tau,q,\ell})+v_q[r,k,m],
\end{aligned}
\label{eq:ofdmreceived}
\end{equation}
where $t_{q,m}=t_{q,0}+mT_s$ with block start $t_{q,0}$ and time-domain noise $v_q[r,k,m]$. The constant $\phi_\ell(t_0)-2\pi\nu_\ell t_0$, carrier-reference phase, and illumination gain are included in the free complex gain $\alpha_{q,\ell}$. The half-wavelength array uses direction cosine $u_{q,\ell}$. Following standard OFDM radar processing~\cite{braun2010}, a unitary discrete Fourier transform (DFT) and elementwise division by the known pilots give
\begin{equation}
y_q[r,n,m]=\frac{1}{p_q[n,m]\sqrt{K_F}}
\sum_{k=0}^{K_F-1}z_q[r,k,m]e^{-j2\pi f_nkT_{\rm samp}}.
\label{eq:pilotremoval}
\end{equation}
Substituting~\eqref{eq:ofdmwaveform} into~\eqref{eq:ofdmreceived} and using subcarrier orthogonality in~\eqref{eq:pilotremoval} gives the CSI model
\begin{subequations}\label{eq:raw}
\begin{align}
y_q[r,n,m]&=\sum_{\ell\in\mathcal L_q}s_{q,\ell}[r,n,m]+w_q[r,n,m],\label{eq:rawsum}\\
s_{q,\ell}[r,n,m]&=\alpha_{q,\ell}e^{j\pi ru_{q,\ell}}
e^{-j2\pi f_n\theta_{\tau,q,\ell}}
e^{j2\pi t_{q,m}\theta_{\nu,q,\ell}}.\label{eq:pathresponse}
\end{align}
\end{subequations}
The indices span $r=0,\ldots,N_r-1$, $n=0,\ldots,N-1$, and $m=0,\ldots,M_q-1$. Independent time-sample noise $v_q\sim\CN(0,\sigma_w^2)$ retains its variance in $w_q$ after the unitary DFT and unit-modulus pilot removal. Gains and arrival directions are unknown in each direction. The geometry check uses interior angles $\vartheta_{q,\ell}$ from the baseline, so $u_{q,\ell}=\cos\vartheta_{q,\ell}$ and $d_{AB}=\|\mathbf b_A-\mathbf b_B\|_2$.
The effective delays lie in the CP window, and we assume front-end synchronization leaves negligible intra-symbol phase variation and sampling dilation over a block~\cite{brunner2025,han2025ota}. Section~\ref{sec:numerical} evaluates clock variation within and between the two blocks.

Stacking with antenna index fastest, subcarrier next, and symbol last gives the separable structure
\begin{equation}
\begin{aligned}
\mathbf y_q=\sum_{\ell\in\mathcal L_q}\alpha_{q,\ell}\bigl[&\mathbf d_q(\theta_{\nu,q,\ell})\otimes\mathbf f(\theta_{\tau,q,\ell})
\\&\otimes\mathbf a_q(u_{q,\ell})\bigr]+\mathbf w_q,
\end{aligned}
\label{eq:vecmodel}
\end{equation}
where $[\mathbf f(\theta)]_n=e^{-j2\pi f_n\theta}$, $[\mathbf d_q(\theta)]_m=e^{j2\pi t_{q,m}\theta}$, and $[\mathbf a_q(u)]_r=e^{j\pi ru}$. The free gain absorbs the constant phase from a change of block origin. Equation~\eqref{eq:vecmodel} supplies both the spatial separator and the raw-observation Jacobian in Section~\ref{sec:theory}.

\subsection{Effective Parameters and Unknown Correspondence}
A single direction identifies only the sums in~\eqref{eq:effective}. A common shift in all physical delays can be exchanged for an opposite timing-origin shift, and the same ambiguity couples Doppler and CFO. For a known common identity, the directional half-sum recovers the physical parameter and the difference recovers twice the offset. With unordered, partially overlapping path lists, the receiver must first identify which differences represent the same inter-node offset. Because each directional block uses one station clock pair, its offset is shared by every visible path.

Uniform sampling makes effective delay and Doppler periodic with periods $1/\Delta f$ and $1/T_s$. The local estimator returns principal phase slopes, and both lists use the same unwrapped branches when compared. In the modeled regime, the physical ranges and clock scales keep the effective slopes within this region. For a correct pair, the nonconjugate product in Section~\ref{sec:recovery} doubles the physical phase slopes, with physical periods $c/(2\Delta f)$ and $1/(2T_s)$, and the receiver selects a branch within the admissible physical interval. Repeated differences between the local effective-parameter lists provide the shared structure for identifying a partial correspondence $\mathcal P$.

\section{Path Association Under a Shared Offset}\label{sec:association}
\begin{figure}[t]
\centering
\includegraphics[width=\columnwidth]{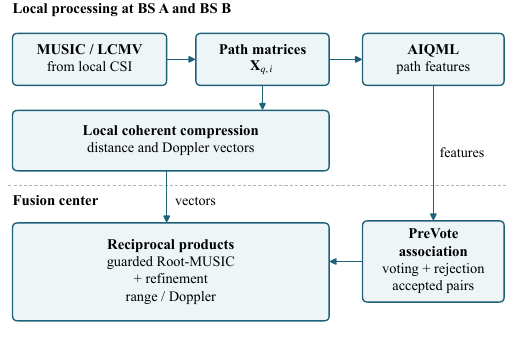}
\caption{Proposed reciprocal receiver: local coherent compression and association-guided fusion at the center.}
\label{fig:receiver_flow}
\end{figure}

Fig.~\ref{fig:receiver_flow} shows how each station turns its array observation into local path descriptors. Before assigning paths, the center tests which trial clock displacement agrees with several pairs of descriptors. The assignment may then leave a path unmatched at either station.

\subsection{Local Slope Estimation and Reliability}
The local receiver converts an array observation into effective delay-Doppler slopes and a measure of their fit quality. Multiple signal classification (MUSIC)~\cite{schmidt1986music} estimates the directional cosines from the forward-backward averaged covariance. A linearly constrained minimum variance (LCMV) separator~\cite{frost1972lcmv} isolates the associated responses. For unfolded CSI $\mathbf Y_q\in\mathbb C^{N_r\times NM_q}$, set $\mathbf R_q=\mathbf Y_q\mathbf Y_q^{\mathrm H}/(NM_q)$ and form $\widehat{\mathbf A}_q$ from the steering vectors at the MUSIC direction estimates. With the loaded covariance $\widetilde{\mathbf R}_q=\mathbf R_q+\epsilon_R\mathbf I$, where $\epsilon_R=\epsilon_{\rm rel}\tr(\mathbf R_q)/N_r+\epsilon_{\rm abs}$ and $\epsilon_G>0$, the LCMV separator is
\begin{equation}\label{eq:lcmvseparator}
\mathbf W_q=\widetilde{\mathbf R}_q^{-1}\widehat{\mathbf A}_q
\bigl(\widehat{\mathbf A}_q^{\mathrm H}\widetilde{\mathbf R}_q^{-1}\widehat{\mathbf A}_q+\epsilon_G\mathbf I\bigr)^{-1}.
\end{equation}
Let $\mathbf w_{q,i}$ be column $i$ of $\mathbf W_q$. Reshaping $\mathbf w_{q,i}^{\mathrm H}\mathbf Y_q$ into $N\times M_q$ gives the local working model
\begin{equation}
\mathbf X_{q,i}=\beta_{q,i}\mathbf a_N(\mu_{q,i})\mathbf b_{M_q}(\omega_{q,i})^{\mathrm{T}}+\mathbf E_{q,i}.
\label{eq:rankone}
\end{equation}
Here $\mathbf a_N(\mu)=[1,e^{j\mu},\ldots,e^{j(N-1)\mu}]^{\mathrm{T}}$ indexes the subcarriers, and $\mathbf b_{M_q}(\omega)=[1,e^{j\omega},\ldots,e^{j(M_q-1)\omega}]^{\mathrm{T}}$ indexes the OFDM symbols. The phase slopes are related to effective delay and Doppler by
\begin{equation}\label{eq:slopemapping}
\mu_{q,i}=-2\pi\Delta f\theta_{\tau,q,i},\qquad
\omega_{q,i}=2\pi T_s\theta_{\nu,q,i}.
\end{equation}
For an isolated path, the centered frequency grid and block origin in~\eqref{eq:raw} contribute only to its complex gain
\begin{equation}
\begin{aligned}
\beta_{q,i}&=(\mathbf w_{q,i}^{\mathrm H}\mathbf a_q(u_{q,i}))\alpha_{q,i}\\
&\quad\times e^{j\pi N\Delta f\theta_{\tau,q,i}}
 e^{j2\pi t_{q,0}\theta_{\nu,q,i}}.
\end{aligned}
\label{eq:localgainorigin}
\end{equation}
With a spatial separator fixed independently of the noise, projected receiver noise is white across samples within that output, with variance $\sigma_w^2\|\mathbf w_{q,i}\|_2^2$. The separated outputs can share receiver noise. The residual $\mathbf E_{q,i}$ also contains leakage from other paths, and the separator is estimated from the observations. We use the squared residual to fit each local rank-one response.

Write $\mathbf S=\mathbf a_N(\mu)\mathbf b_{M_q}(\omega)^{\mathrm{T}}$. Profiling out the complex gain gives $\widehat\beta=\langle\mathbf S,\mathbf X\rangle_F/(NM_q)$, or explicitly,
\begin{equation}
\widehat\beta(\mu,\omega)=\frac{\mathbf a_N(\mu)^{\mathrm H}\mathbf X\mathbf b_{M_q}(\omega)^*}{NM_q},
\label{eq:aiqmlgain}
\end{equation}
where the local indices are suppressed. Substituting~\eqref{eq:aiqmlgain} into $\|\mathbf X-\beta\mathbf S\|_F^2$ eliminates the gain and yields the profiled objective
\begin{equation}
\mathcal L(\mu,\omega)
=\|\mathbf X\|_F^2-
\frac{|\mathbf a_N(\mu)^{\mathrm H}\mathbf X\mathbf b_{M_q}(\omega)^*|^2}{NM_q}.
\label{eq:aiqmlobjective}
\end{equation}
The profiled objective is minimized at a maximum of the two-dimensional periodogram, which supplies the initial slopes. We evaluate it using an $N_f\times N_t$ fast Fourier transform (FFT) with $N_f=2^{\lceil\log_2N\rceil}$ and $N_t=\max\{1024,2^{\lceil\log_2M_q\rceil}\}$. Each maximizing bin is mapped to its principal phase.

The subsequent updates use approximate iterative quadratic maximum-likelihood (AIQML) refinement~\cite{so2006}. For each axis, use $\mathbf Z\in\mathbb C^{D\times K}$ equal to $\mathbf X$ or $\mathbf X^{\mathrm{T}}$. Define adjacent slices $\mathbf Z_-=[\mathbf I_{D-1}\ \mathbf0]\mathbf Z$ and $\mathbf Z_+=[\mathbf0\ \mathbf I_{D-1}]\mathbf Z$. The residual $\mathbf R_D(\vartheta)=\mathbf Z_+-e^{j\vartheta}\mathbf Z_-$ uses shared samples in adjacent differences. For white input noise, its normalized covariance has main-diagonal entries $2$ and first subdiagonal entries $-e^{j\vartheta}$. The resulting inverse is
\begin{equation}\label{eq:aiqmlweight}
[\mathbf V_D(\vartheta)]_{a,b}=\left(\min(a,b)-\frac{ab}{D}\right)e^{j(a-b)\vartheta},\quad 1\leq a,b<D.
\end{equation}

Fixing the weight at the current slope $\vartheta^{(t)}$ leaves a single phase term in the objective
\begin{equation}\label{eq:aiqmlquadratic}
\begin{aligned}
Q(\vartheta\mid\vartheta^{(t)})
&=\tr\{\mathbf R_D(\vartheta)^{\mathrm H}\mathbf V_D(\vartheta^{(t)})\mathbf R_D(\vartheta)\}\\
&=C_t-2\Real\{e^{-j\vartheta}\tr(\mathbf Z_-^{\mathrm H}\mathbf V_D(\vartheta^{(t)})\mathbf Z_+)\},
\end{aligned}
\end{equation}
where $C_t$ is independent of phase. Aligning the trial phase with the complex cross term gives
\begin{equation}\label{eq:aiqmlupdate}
\vartheta^{(t+1)}=\arg\tr\!\left(\mathbf Z_-^{\mathrm H}\mathbf V_D(\vartheta^{(t)})\mathbf Z_+\right).
\end{equation}
Each sweep updates the frequency slope and then the symbol slope, using the ordinary transpose for the second update. Cumulative sums apply the structured weight in $O(DK)$ per axis, giving $O(I_aNM_q)$ for $I_a$ sweeps after initialization. The profiled objective~\eqref{eq:aiqmlobjective} supplies both initialization and fit quality. Root mean square (RMS) normalization preserves phase slopes and the normalized residual $e=\mathcal L/\|\mathbf X\|_F^2$. Each path sends five real features comprising angle, effective delay, effective Doppler, original log power $p=\log(\|\mathbf X\|_F^2/(NM_q))$, and inverse residual $r=1/\max(e,\epsilon_0)$, with numerical floors for zero energy. The center sets $a=\log(1+r)$ and $Q=\operatorname{clip}_{[q_0,1]}\{a/(a+a_{\rm med}+\epsilon_0)\}$, where $a_{\rm med}$ is the median positive local value and $\epsilon_0$ is machine precision. Nonfinite or negative raw reliabilities are set to zero before mapping, and a list without positive values receives unit scores.

\subsection{Shared Displacement and Compatibility}
Correct reciprocal paths share a translation in the effective delay-Doppler plane. For pair $(i,j)$ define $\boldsymbol\Delta_{ij}=(\widehat\theta_{\tau,A,i}-\widehat\theta_{\tau,B,j},\widehat\theta_{\nu,A,i}-\widehat\theta_{\nu,B,j})^{\mathrm{T}}$ and $\boldsymbol\delta=(2\tau_o,2f_o)^{\mathrm{T}}$. If the entries originate from physical paths $a$ and $b$, then
\begin{equation}\label{eq:displacementerror}
\boldsymbol\Delta_{ij}=\boldsymbol\delta+
\begin{bmatrix}\tau_a-\tau_b\\\nu_a-\nu_b\end{bmatrix}
+\boldsymbol\varepsilon_{A,i}-\boldsymbol\varepsilon_{B,j},
\end{equation}
where $\boldsymbol\varepsilon_{q,i}=(\widehat\theta_{\tau,q,i}-\theta_{\tau,q,i},\widehat\theta_{\nu,q,i}-\theta_{\nu,q,i})^{\mathrm{T}}$ is the local estimation-error vector. The errors may be biased. A correct pair has zero physical delay and Doppler differences, so every such pair supports the same clock displacement in the absence of estimation error. Pairing a private path with an unrelated response can support a different displacement. Independent centering incorporates the medians of the two visible sets. If $m_{x,q}$ is the physical median of coordinate $x\in\{\tau,\nu\}$ in visible set $\mathcal L_q$, exact local parameters satisfy
\begin{equation}\label{eq:centeringbias}
\operatorname{med}_i\theta_{x,A,i}-\operatorname{med}_j\theta_{x,B,j}
=\delta_x+m_{x,A}-m_{x,B}.
\end{equation}
The offset inferred by centering thus includes a term set by directional visibility. We therefore search repeated pairwise displacements in $\mathcal H=\{\mathbf h_0\}\cup\{\boldsymbol\Delta_{ij}\}$, where $\mathbf h_0$ is the coordinate-wise difference of local medians.

Compatibility combines this shared-offset relation with evidence available for each candidate identity. Let $q_{ij}=\sqrt{Q_{A,i}Q_{B,j}}$, $\rho_{x,ij}(\boldsymbol\delta)=|\Delta_{x,ij}-\delta_x|/s_x$, and $\rho_{p,ij}=|p_{A,i}-p_{B,j}|/s_p$. The scales use the Gaussian-consistent median absolute deviation (MAD) of the separately centered local lists. A zero or nonfinite MAD is replaced by sample standard deviation before enforcing $s_\tau\geq b_\tau$, $s_\nu\geq b_\nu$, and $s_p\geq s_{p,\min}$. The physical half-sums $\bar\theta_{x,ij}=(\widehat\theta_{x,A,i}+\widehat\theta_{x,B,j})/2$ define a plausibility event $\mathcal V_{ij}$ through public delay and Doppler intervals. Displacement agreement is $\mathcal W_{ij}(\boldsymbol\delta)=\{|\Delta_{\tau,ij}-\delta_\tau|\leq b_\tau,\ |\Delta_{\nu,ij}-\delta_\nu|\leq b_\nu\}$. For a common scatterer, applying the sine rule to the baseline and the two interior look angles gives the bistatic path length
\begin{equation}\label{eq:geomgate}
R_{ij}^{\rm ang}=d_{AB}\frac{\sin\vartheta_{A,i}+\sin\vartheta_{B,j}}{\sin(\vartheta_{A,i}+\vartheta_{B,j})}.
\end{equation}
For the exact parameters of a common path, $R_{ij}^{\rm del}=c\bar\theta_{\tau,ij}=R_{ij}^{\rm ang}$ because the timing offsets in~\eqref{eq:effective} cancel in the half-sum. The difference $r_{g,ij}=|R_{ij}^{\rm del}-R_{ij}^{\rm ang}|$ measures violation of this condition. Since it uses a clock-free range, this check helps distinguish candidates that support the same trial displacement. For a well-conditioned intersection with admissible range, $G_{ij}=w_gr_{g,ij}/s_g+\lambda_g\mathbf1\{r_{g,ij}>g_{\max}\}$, and otherwise $G_{ij}=0$.

These relations lead to the normalized compatibility cost
\begin{equation}\label{eq:cost}
\begin{aligned}
C_{ij}(\boldsymbol\delta)&=[1+\lambda_r(1-q_{ij})]
[\rho_{\tau,ij}+\rho_{\nu,ij}+w_p\rho_{p,ij}]\\
&\quad+\lambda_Q|Q_{A,i}-Q_{B,j}|+G_{ij}\\
&\quad+\lambda_p\mathbf1_{\mathcal V_{ij}^c}
+\lambda_b\mathbf1_{\mathcal W_{ij}^c(\boldsymbol\delta)}.
\end{aligned}
\end{equation}
Only the slope residuals and their gate in~\eqref{eq:cost} change with the trial displacement. The power, fit, and geometry terms can be evaluated once per candidate pair. At this stage two low-cost candidates may use the same local path. Their votes carry evidence about the clock even though a one-to-one assignment can retain at most one of them.

\subsection{Support Selection and Partial Matching}
By~\eqref{eq:displacementerror}, candidate pairs with similar differences support the same clock displacement. PreVote accumulates that support before imposing a one-to-one correspondence. Let $C^{(0)}$ denote~\eqref{eq:cost} without its displacement penalty. A pair with $e_{ij}=\mathbf1\{C_{ij}^{(0)}(\mathbf h_0)<\tfrac12\min(\lambda_p,\lambda_g)\}$ contributes weight $w_{ij}=e_{ij}\max(q_{ij},q_0)$. The selected displacement maximizes the total weight within its window,
\begin{equation}\label{eq:prevote}
V(\mathbf h)=\sum_{i,j}w_{ij}\mathbf1_{\mathcal W_{ij}(\mathbf h)},
\qquad \widehat{\boldsymbol\delta}_v\in\argmax_{\mathbf h\in\mathcal H}V(\mathbf h).
\end{equation}
A single slope difference fits its own displacement, so repeated evidence requires at least two common paths. To see when their votes identify the correct group, measure distance by $\|\mathbf d\|_{b,\infty}=\max_x |d_x|/b_x$. Suppose the eligible correct differences have diameter at most one and are more than two units from every eligible wrong difference. A window centered on any correct difference then contains the entire correct group, and no window can mix the two groups. If the correct group outweighs every window of wrong differences,~\eqref{eq:prevote} selects a window containing all its members. At the selected displacement, let $\mathfrak M$ contain the disjoint pairings, including the empty pairing. Accounting for the cost of leaving a response unmatched gives
\begin{equation}\label{eq:assign}
\widehat{\mathcal P}\in\argmin_{\mathcal P\in\mathfrak M}
\left\{\sum_{(i,j)\in\mathcal P}C_{ij}(\widehat{\boldsymbol\delta}_v)
+\frac\gamma2(L_A+L_B-2|\mathcal P|)\right\}.
\end{equation}
Rejecting both endpoints of a pair costs $\gamma$, so accepting the pair saves $\gamma-C_{ij}$ relative to rejecting it. If each true pair costs less than $\gamma$ and every other pair costs more, the disjoint true pairs together form the unique minimizer of~\eqref{eq:assign}. The receiver then retains the largest group of assigned pairs consistent with one displacement. Each assigned difference seeds a window, whose center is refined by reliability-weighted coordinate medians. Ties favor greater support weight and then smaller weighted residual. With multiple tentative pairs, a group needs at least two members, while a sole assignment supplies its own center. If the true differences also have diameter at most one in $\|\cdot\|_{b,\infty}$, every member seeds a window containing them all. The median lies within their coordinate bounds, so recentering preserves this group.

\input{expanded_matrix}
\section{Reciprocal Parameter Recovery}\label{sec:recovery}
The half-sum of the local slopes gives a direct way to see what a reciprocal product measures. The coherent receiver estimates this quantity from the phase samples supplied by the two stations.

\subsection{Reciprocal Cancellation and Pairing Errors}
For an accepted pair $(i,j)$, the scalar estimates are
\begin{equation}\label{eq:scalarfusion}
\widehat R^{\rm sc}_{ij}=\frac c2(\widehat\theta_{\tau,A,i}+\widehat\theta_{\tau,B,j}),\qquad
\widehat\nu^{\rm sc}_{ij}=\frac12(\widehat\theta_{\nu,A,i}+\widehat\theta_{\nu,B,j}).
\end{equation}
Substituting~\eqref{eq:effective} cancels the opposite constant offsets. For a correct pair the half-sum then estimates one physical coordinate. If the local errors in either coordinate are $\varepsilon_A,\varepsilon_B$, let $b_q=\E\varepsilon_q$ denote their biases for $q\in\{A,B\}$. The physical-coordinate error is $\bar\varepsilon=(\varepsilon_A+\varepsilon_B)/2$, with bias $b_{\rm sc}=(b_A+b_B)/2$. Writing $\sigma_q^2=\operatorname{var}(\varepsilon_q)$ and $\sigma_{AB}=\operatorname{cov}(\varepsilon_A,\varepsilon_B)$ gives
\begin{equation}
\operatorname{var}(\bar\varepsilon)=\tfrac14\bigl[\sigma_A^2+\sigma_B^2+2\sigma_{AB}\bigr].
\label{eq:halfsumerror}
\end{equation}
Distance error is $c\bar\varepsilon_\tau$. Each slope uses all $M_q$ local symbols, even for unequal blocks. Time-varying clocks add the residual analyzed in Section~\ref{sec:theory}.

With exact local effective parameters, constant antisymmetric offsets, and consistent phase branches, a pair formed from distinct targets $a$ and $b$ returns
\begin{equation}\label{eq:wrongpair}
\widehat R_{ab}=(R_a+R_b)/2,\qquad \widehat\nu_{ab}=(\nu_a+\nu_b)/2.
\end{equation}
Substituting $(\tau_a+\tau_o,\tau_b-\tau_o)$ into the half-sum gives error $(R_b-R_a)/2$ relative to target $a$. Doppler similarly gives $(\nu_b-\nu_a)/2$. Since a midpoint can lie near either path, the experiments score physical accuracy and correct identity separately.

\ReserveExpandedMatrixNumber
For a reciprocal pair $(i,j)$ associated with path $\ell$, the local observation at each station has the matrix form in~\eqref{eq:expandedmatrix}, shown at the bottom of this page. Here $i_{+1}=i$, $i_{-1}=j$, and $M=M_q$. Variation with subcarrier index $n$ encodes $\tau_\ell+q\tau_o$, and variation with symbol index $m$ encodes $\nu_\ell+qf_o$. The gain $\beta_{q,i_q}$ includes the origin phases derived in~\eqref{eq:localgainorigin}.

\subsection{Coherent Compression and Reciprocal Fusion}
Each node compresses its local path observations while their effective slopes still contain the clock terms. A reference formed from the central window sets one phase correction per column. Power weights then combine the corrected columns along the retained phase axis. The center multiplies the compressed observations of paths selected by PreVote and estimates each physical slope from the resulting phase samples.

For any nonzero-energy $\mathbf Z\in\mathbb C^{D\times K}$, let $s_Z=(\|\mathbf Z\|_F^2/(DK))^{1/2}$ and $\widetilde{\mathbf Z}=\mathbf Z/s_Z$. With zero-based center $d_c=\lfloor D/2\rfloor$, let $\mathcal I_c=\{\max(0,d_c-2),\ldots,\min(D-1,d_c+2)\}$. Its indicator vector $\mathbf v_c\in\{0,1\}^D$ defines
\begin{equation}\label{eq:phasereference}
\mathbf r^{\mathrm{T}}=\mathbf v_c^{\mathrm{T}}\widetilde{\mathbf Z},\qquad \phi_k=e^{-j\arg r_k}.
\end{equation}
Column powers $p_k=\|\widetilde{\mathbf Z}_{:,k}\|_2^2/D$ give normalized weights $w_k=p_k/\sum_a p_a$. If $|r_k|\leq\epsilon_{\rm ref}\max\{\sum_{d\in\mathcal I_c}|\widetilde Z[d,k]|,\epsilon_0\}$, the largest-magnitude sample in that column's window replaces $r_k$, with the first row breaking ties. A zero reference has zero phase, and a zero-energy column has zero weight. With $\mathbf Z_{\rm cal}=\widetilde{\mathbf Z}\diag(\boldsymbol\phi)$, phase alignment and weighted accumulation give
\begin{equation}\label{eq:compression}
\mathcal C(\mathbf Z)=\mathbf Z_{\rm cal}\mathbf w=\mathbf Z\mathbf c(\mathbf Z),\qquad
\mathbf c(\mathbf Z)=\frac{\boldsymbol\phi\odot\mathbf w}{s_Z}.
\end{equation}
The unit-modulus factors $\phi_k$ align columns without changing relative amplitudes. Data-dependent phases, RMS scaling, and weights make $\mathcal C$ generally nonlinear, with potentially correlated and biased errors.

\begin{lemma}[Ideal slope preservation]\label{lem:slope}
In exact arithmetic, for $\mathbf Z=\beta\mathbf a\mathbf b^{\mathrm{T}}$ with $\beta\ne0$ and unit-modulus entries of $\mathbf a,\mathbf b$, compression gives $\mathcal C(\mathbf Z)=C_a\mathbf a$ for a nonzero constant $C_a$.
\end{lemma}
\begin{proof}[\hspace*{\parindent}Proof]
Let $A_c=\sum_{d\in\mathcal I_c}a_d$. Then $s_Z=|\beta|$, $r_k=(\beta/|\beta|)A_cb_k$, and $w_k=1/K$. With the summed reference, every calibrated column equals $e^{-j\arg A_c}\mathbf a$. If $A_c$ is zero or triggers the fallback, the unit-modulus entries tie in magnitude and every column selects the first window row. Its $a_d$ replaces $A_c$, so the weighted sum still has the same slope.
\end{proof}
By \textbf{Lemma~\ref{lem:slope}}, compression changes only the complex scale of a rank-one signal term. For measured responses the coefficients in~\eqref{eq:compression} also depend on the observed noise.

To form the range and Doppler vectors, suppress $i_q$ and let $\mathbf X_q=\mathbf X_{q,i_q}$, $M_c=\min(M_A,M_B)$, and define
$\mathbf X_q^{(c)}$ as the first $M_c$ columns of $\mathbf X_q$. The two measurement vectors are
\begin{equation}\label{eq:truncatefirst}
\mathbf g_{R,q}=\mathcal C(\mathbf X_q),\qquad
\mathbf g_{\nu,q}=\mathcal C\!\left((\mathbf X_q^{(c)})^{\mathrm{T}}\right).
\end{equation}
The length-$N$ range vector uses all $M_q$ columns. For the length-$M_c$ Doppler vector, truncation precedes ordinary transposition, normalization, reference selection, and weighting. Thus unequal blocks restrict its common aperture, while ordinary transposition exchanges the phase axes without conjugation.

With the retained slopes established, the two directional vectors can be combined. For a correct pair, their ideal nonconjugate products satisfy
\begin{equation}\label{eq:product}
h_R[n]=A_Re^{-j4\pi n\Delta f\tau_\ell},\qquad
h_\nu[m]=A_\nu e^{j4\pi mT_s\nu_\ell}.
\end{equation}
The common subcarrier spacing and symbol period let the products add the slopes separately on the two axes. By~\eqref{eq:effective}, the delay axis uses $(\tau_\ell+\tau_o)+(\tau_\ell-\tau_o)=2\tau_\ell$, and the Doppler axis uses $(\nu_\ell+f_o)+(\nu_\ell-f_o)=2\nu_\ell$. Block-start phases enter $A_R,A_\nu$, and a wrong pair returns the midpoint in~\eqref{eq:wrongpair}.

For ideal entries $s_A,s_B\ne0$ measured as $g_A=s_A+e_A$ and $g_B=s_B+e_B$, the product error is $e_h=s_Ae_B+s_Be_A+e_Ae_B$. If $e_A,e_B$ are independent and zero mean with variances $\sigma_A^2,\sigma_B^2$, then
\begin{equation}
\frac{\E|e_h|^2}{|s_As_B|^2}=\rho_A^{-1}+\rho_B^{-1}+(\rho_A\rho_B)^{-1}.
\label{eq:normalizednoise}
\end{equation}
Here $\rho_q=|s_q|^2/\sigma_q^2$. Equation~\eqref{eq:normalizednoise} isolates a product penalty minimized by balanced entry SNRs at fixed $\rho_A+\rho_B$. The phases and weights in~\eqref{eq:compression} also shape the errors in the measured vectors.

For a finite nonzero product of length $D\geq6$, normalized as $\widetilde{\mathbf h}$, form $H_{a,b}=\widetilde h[a+b]$ with $L_h=\lfloor D/2\rfloor$ rows and $D-L_h+1$ sliding columns. The principal eigenvector $\mathbf u_1$ of $\mathbf H\mathbf H^{\mathrm H}$ defines the noise projector $\mathbf P_n=\mathbf I-\mathbf u_1\mathbf u_1^{\mathrm H}$. Writing $c_k=\sum_{a=0}^{L_h-1-k}[P_n]_{a,a+k}$ for $k\geq0$ and $c_{-k}=c_k^*$, Root-MUSIC finds the roots of $\sum_{k=-(L_h-1)}^{L_h-1}c_kz^{k+L_h-1}$. With $z=e^{j\omega}$, the nonzero root on or inside the unit circle (up to numerical tolerance) and closest to it supplies a principal phase $\arg z$. The subcarrier product yields the estimated phase increment $\widehat\omega_R$ between adjacent tones, and the slow-time product yields $\widehat\omega_\nu$ between adjacent symbols. By~\eqref{eq:product}, the corresponding range and Doppler estimates are
\begin{equation}\label{eq:recovered}
\widehat R_\ell=-c\widehat\omega_R/(4\pi\Delta f),\quad
\widehat\nu_\ell=\widehat\omega_\nu/(4\pi T_s).
\end{equation}

\subsection{Communication and Computational Cost}
Let $b_r$ and $b_c$ be the bytes per real and complex entry. Scalar features require $B_{\rm sc}=5b_r(L_A+L_B)$ bytes. Raw CSI requires $B_{\rm raw}=b_cN_rN(M_A+M_B)$, and separated matrices require $B_{\rm path}=b_cN(L_AM_A+L_BM_B)$. Coherent vectors for both local path lists require $B_{\rm vec}=b_c(L_A+L_B)(N+M_c)$ bytes. The receiver transmits the descriptors and both vectors for every local path. For fixed identities and phase branches, the payload is a statistic of the raw measurements, so its Fisher information is bounded by that of those measurements. With double-precision entries ($b_r=8$, $b_c=16$), the descriptors and compressed vectors respectively occupy 240 and 44\,160 bytes in S1, compared with 25\,804\,800 bytes for raw CSI. The corresponding S2 payloads are 320 and 58\,880 bytes. The budgets count all local paths and exclude indices and framing. The center forms products for the selected pairs. Requesting vectors only for the $P$ accepted pairs would send $2b_cP(N+M_c)$ vector bytes and require feedback of the accepted indices and a second transmission round. The controls share MUSIC/LCMV/AIQML processing. Writing $H=|\mathcal H|\leq1+L_AL_B$, cost construction is $O(L_AL_B)$ and voting is $O(HL_AL_B)$. Padded assignment costs $O((L_A+L_B)^3)$. For $P_0=|\widehat{\mathcal P}|\geq2$ tentative pairs, consensus tests $P_0$ seeds with at most $I_c=5$ updates and sorted weighted medians, costing $O(I_cP_0^2\log P_0)$. Full-search repeats assignment across hypotheses and iterations. AIQML costs $O(I_aNM_q)$ per path. Compression scans retained entries.

\section{Identifiability and Estimation Bounds}\label{sec:theory}
A common path supplies two observations of one physical delay and Doppler, with clock offsets of opposite sign. We first ask when these observations identify the clock and then quantify the cost of estimating it from the same data.

\subsection{Identifiability With Reciprocal Observations}
For one physical axis, write $x_\ell\in\{\tau_\ell,\nu_\ell\}$ and $b\in\{\tau_o,f_o\}$. With known cross-direction identities, stack the common physical parameters as $\mathbf x_c=(x_\ell)_{\ell\in\mathcal L_c}$ and the corresponding directional effective parameters into $\mathbf z$. They satisfy
\begin{equation}
\mathbf z=\begin{bmatrix}\mathbf I&\mathbf 1\\\mathbf I&-\mathbf 1\end{bmatrix}
\begin{bmatrix}\mathbf x_c\\b\end{bmatrix}.
\label{eq:idmatrix}
\end{equation}
For $L_c\geq1$, the null space is trivial since $x_\ell=(z_{+,\ell}+z_{-,\ell})/2$ and $b=(z_{+,\ell}-z_{-,\ell})/2$. One correctly identified reciprocal path anchors the clock, permitting one-sided recovery $x_\ell=z_{q,\ell}-qb$. Without a common path or equivalent external constraint, physical parameters absorb the clock shift.

With unknown path identities and fixed phase branches, let $\mathbf h=(h_\tau,h_\nu)^{\mathrm{T}}$ denote a candidate displacement in effective delay (seconds) and Doppler (hertz). A pair belongs to $\mathcal P(\mathbf h)$ if it is admissible and its BS A parameter vector equals its BS B vector plus $\mathbf h$. For a common path the exact displacement is $\boldsymbol\delta=(2\tau_o,2f_o)^{\mathrm{T}}$ by~\eqref{eq:displacementerror}. Distinct parameter vectors within each list give each response at most one counterpart, so these pairs form the largest matching at $\mathbf h$. If only one displacement attains a positive maximum of $|\mathcal P(\mathbf h)|$, it specifies a unique maximum-overlap matching with implied clock vector $\mathbf h/2$. Tied maxima give different matchings of equal size. The physical and geometric constraints in Section~\ref{sec:association} can distinguish candidates left ambiguous by the slopes.

For the real Jacobian defined below, let $\mathcal D_x$ correspond to physical coordinates $\mathbf x=(\boldsymbol\tau^{\mathrm{T}},\boldsymbol\nu^{\mathrm{T}})^{\mathrm{T}}$ and $\mathcal D_\zeta=[\mathcal D_\gamma,\mathcal D_u,\mathcal D_b]$ to the complex gains, directional cosines $\mathbf u$, and clock vector $\mathbf b=(\tau_o,f_o)^{\mathrm{T}}$. The physical coordinates are distinguishable from nuisance perturbations to first order when
\begin{equation}
\operatorname{rank}([\mathcal D_\zeta,\mathcal D_x])
-\operatorname{rank}(\mathcal D_\zeta)=\dim(\mathbf x).
\label{eq:rankcondition}
\end{equation}
This rank test uses fixed identities and phase branches after any gauge constraints. For an unwrapped coordinate, an admissible interval narrower than its alias period contains at most one representative of a given phase.

\subsection{Information in the Array Observations and Nuisance Elimination}
Let $L=|\mathcal L_A\cup\mathcal L_B|$, $V=L_A+L_B$, and $K=N_rN(M_A+M_B)$. The real parameter vector $\boldsymbol\eta=(\boldsymbol\tau^{\mathrm{T}},\boldsymbol\nu^{\mathrm{T}},\mathbf b^{\mathrm{T}},\mathbf u^{\mathrm{T}},\boldsymbol\gamma^{\mathrm{T}})^{\mathrm{T}}$ has dimension $2L+2+3V$, with $\boldsymbol\gamma=(\Real\boldsymbol\alpha^{\mathrm{T}},\operatorname{Im}\boldsymbol\alpha^{\mathrm{T}})^{\mathrm{T}}$. Each visible path has a free directional angle and complex gain. The information model treats angles and delays as independent coordinates. The receiver uses their geometric relation only for association. We evaluate the information at each realization while treating channel and clock quantities as unknown deterministic parameters. For $\mathbf y\sim\CN(\boldsymbol\mu(\boldsymbol\eta),\sigma_w^2\mathbf I)$ and $\mathbf G=\partial\boldsymbol\mu/\partial\boldsymbol\eta^{\mathrm{T}}$, the noise-whitened real Jacobian is
\begin{equation}
\mathcal D=\frac{\sqrt{2}}{\sigma_w}
\begin{bmatrix}\Real\mathbf G\\\operatorname{Im}\mathbf G\end{bmatrix}
\in\mathbb R^{2K\times(2L+2+3V)}.
\label{eq:realjacobian}
\end{equation}
The Fisher information matrix (FIM) is $\mathbf J=\mathcal D^{\mathrm{T}}\mathcal D$, or elementwise,
\begin{equation}
J_{ij}=\frac{2}{\sigma_w^2}\Real\left\{
\left(\frac{\partial\boldsymbol\mu}{\partial\eta_i}\right)^{\mathrm H}
\frac{\partial\boldsymbol\mu}{\partial\eta_j}\right\}.
\label{eq:fim}
\end{equation}
Differentiating~\eqref{eq:pathresponse} multiplies a path response by $-j2\pi f_n$, $j2\pi t_{q,m}$, or $j\pi r$ for delay, Doppler, or direction cosine. The two gain derivatives are its unit-gain response and $j$ times that response. Since every visible path shares the same signed offsets,
\begin{equation}\label{eq:sharedderivative}
\partial_{\tau_o}\mu_q=-j2\pi qf_n\sum_{\ell\in\mathcal L_q}s_{q,\ell},\quad
\partial_{f_o}\mu_q=j2\pi qt_{q,m}\sum_{\ell\in\mathcal L_q}s_{q,\ell}.
\end{equation}
The inner products in~\eqref{eq:fim} retain both cross-path terms and the common clock columns.

Partition $\boldsymbol\eta$ into the delay and Doppler coordinates $\mathbf x=(\boldsymbol\tau^{\mathrm{T}},\boldsymbol\nu^{\mathrm{T}})^{\mathrm{T}}$ and the nuisance coordinates $\boldsymbol\zeta=(\boldsymbol\gamma^{\mathrm{T}},\mathbf u^{\mathrm{T}},\mathbf b^{\mathrm{T}})^{\mathrm{T}}$. Projecting the physical derivatives away from the nuisance span gives the effective information. With nonsingular nuisance and effective blocks, its inverse gives the joint bound
\begin{equation}\label{eq:schur}
\mathbf J_x^e=\mathbf J_{xx}-\mathbf J_{x\zeta}\mathbf J_{\zeta\zeta}^{-1}\mathbf J_{\zeta x},\qquad
\mathbf B_x=(\mathbf J_x^e)^{-1}.
\end{equation}
The full information matrix is evaluated at a common parameter point before projection. The elimination removes gain derivatives, then arrival angle derivatives, then the shared clock directions. For $\boldsymbol\varphi=(\mathbf x^{\mathrm{T}},\mathbf b^{\mathrm{T}})^{\mathrm{T}}$ and $a,c\in\{u,\varphi\}$, the corresponding Schur complements are
\begin{subequations}\label{eq:stagedschur}
\begin{align}
\mathbf J^{(g)}_{ac}
&=\mathbf J_{ac}-\mathbf J_{a\gamma}\mathbf J_{\gamma\gamma}^{-1}\mathbf J_{\gamma c},\\
\mathbf J^{(g,u)}_{\varphi\varphi}
&=\mathbf J^{(g)}_{\varphi\varphi}-\mathbf J^{(g)}_{\varphi u}
(\mathbf J^{(g)}_{uu})^{-1}\mathbf J^{(g)}_{u\varphi},\\
\mathbf J_x^e
&=\mathbf J^{(g,u)}_{xx}-\mathbf J^{(g,u)}_{xb}
(\mathbf J^{(g,u)}_{bb})^{-1}\mathbf J^{(g,u)}_{bx}.
\end{align}
\end{subequations}
Updating each remaining cross block makes the staged elimination equivalent to~\eqref{eq:schur}. Let $\mathbf P_\zeta^\perp$ project off the nuisance span. The mixed block $\mathbf J_{\tau\nu}^e=\mathcal D_\tau^{\mathrm{T}}\mathbf P_\zeta^\perp\mathcal D_\nu$ can remain nonzero after projection. The marginal delay bound is the inverse of $\mathbf J_{\tau\tau}^e-\mathbf J_{\tau\nu}^e(\mathbf J_{\nu\nu}^e)^{-1}\mathbf J_{\nu\tau}^e$, and exchanging the axes gives the Doppler bound. Section~\ref{sec:clock_covariance} identifies conditions that make the mixed blocks vanish. For evaluation, column scaling stabilizes inversion, with singular bounds recorded as unavailable. Range covariance is $c^2$ times delay covariance.

For an externally fixed identity set $\mathcal P$ and ambiguity branch $\mathcal B$, the corresponding CRB for locally unbiased estimation is
\begin{equation}
\operatorname{Cov}_{\rm fixed}(\widehat{\mathbf x}\mid\mathcal P,\mathcal B)\succeq\mathbf B_x.
\label{eq:conditionalcov}
\end{equation}
Let $\mathbf F$, $\mathbf C$, and $\mathbf H$ denote the physical, cross, and clock blocks after eliminating gains and angles. For a positive-definite joint information matrix on these coordinates, the clock Schur complement is $\mathbf S_b=\mathbf H-\mathbf C^{\mathrm{T}}\mathbf F^{-1}\mathbf C$. Block inversion gives the increment in the covariance bound $\mathbf B_x-\mathbf F^{-1}=\mathbf F^{-1}\mathbf C\mathbf S_b^{-1}\mathbf C^{\mathrm{T}}\mathbf F^{-1}\succeq\mathbf 0$. Since $\mathbf C$ has two columns, this clock penalty has rank at most two even with coupled paths and axes. A linear contrast $\mathbf a^{\mathrm{T}}\mathbf x$ with $\mathbf C^{\mathrm{T}}\mathbf F^{-1}\mathbf a=\mathbf 0$ retains the known-clock variance bound. The raw-CSI benchmark in Section~\ref{sec:numerical} uses the diagonal entries of $\mathbf B_x$, with delay variances scaled by $c^2$ for range.

\subsection{Shared-Clock Covariance}\label{sec:clock_covariance}
For known angles and isolated paths with orthogonal response/derivative subspaces, eliminating a free complex gain projects with $\mathbf P_v^\perp=\mathbf I-\mathbf v\mathbf v^{\mathrm H}/(\mathbf v^{\mathrm H}\mathbf v)$. On the rectangular constant-magnitude grid this centers $f_n$ and $t_{q,m}$. Suppressing $[r,n,m]$ gives
\begin{equation}\label{eq:centeredderivatives}
\begin{aligned}
\widetilde{\partial_{\theta_\tau}s}_{q,\ell}&=-j2\pi(f_n-\bar f)s_{q,\ell},\\
\widetilde{\partial_{\theta_\nu}s}_{q,\ell}&=j2\pi(t_{q,m}-\bar t_q)s_{q,\ell}.
\end{aligned}
\end{equation}
The delay and Doppler directions are orthogonal because $\sum_{n,m}(f_n-\bar f)(t_{q,m}-\bar t_q)=0$. The TO and CFO derivatives lie in the respective subspaces, so their elimination preserves this separation. Uneven weights can retain cross terms. Substitution into~\eqref{eq:fim} gives
\begin{equation}\label{eq:lambda}
\lambda_{\tau,q,\ell}=8\pi^2\rho_{q,\ell}M_qS_f,\qquad
\lambda_{\nu,q,\ell}=8\pi^2\rho_{q,\ell}NS_{t,q},
\end{equation}
where $\rho_{q,\ell}=N_r|\alpha_{q,\ell}|^2/\sigma_w^2$, $S_f=\sum_n(f_n-\bar f)^2$, and $S_{t,q}=\sum_m(t_{q,m}-\bar t_q)^2$. The associated local bounds are $B_{\theta_\tau,q,\ell}=\lambda_{\tau,q,\ell}^{-1}$ and $B_{\theta_\nu,q,\ell}=\lambda_{\nu,q,\ell}^{-1}$.
For uniform sampling, the moments reduce to
\begin{equation}\label{eq:uniformmoments}
S_f=\Delta f^2N(N^2-1)/12,\qquad
S_{t,q}=T_s^2M_q(M_q^2-1)/12.
\end{equation}
These bounds show the separate contributions of the frequency and slow-time apertures, whereas~\eqref{eq:schur} additionally retains cross-path and angle coupling.

For either axis, $\theta_{\pm,\ell}=x_\ell\pm b$ transforms the local information into the pair block $\left[\begin{smallmatrix}d_\ell&g_\ell\\g_\ell&d_\ell\end{smallmatrix}\right]$, where $d_\ell=\lambda_{+,\ell}+\lambda_{-,\ell}>0$ and $g_\ell=\lambda_{+,\ell}-\lambda_{-,\ell}$. A missing direction has zero information. The coordinates share one clock $b$, so
\begin{equation}
\mathbf J_{x,b}=\begin{bmatrix}\mathbf D&\mathbf g\\\mathbf g^{\mathrm{T}}&h\end{bmatrix},
\quad \mathbf D=\diag(d_1,\ldots,d_L),
\label{eq:kappa}
\end{equation}
with $\mathbf g=(g_1,\ldots,g_L)^{\mathrm{T}}$ and $h=\sum_\ell d_\ell$.
Eliminating the physical coordinates gives the effective information for the shared clock
\begin{equation}
\kappa=h-\mathbf g^{\mathrm{T}}\mathbf D^{-1}\mathbf g
=\sum_\ell\frac{4\lambda_{+,\ell}\lambda_{-,\ell}}{d_\ell}.
\label{eq:clockschur}
\end{equation}
For $\kappa>0$, block inversion then yields
\begin{equation}\label{eq:clockbound}
B_b=\kappa^{-1},\qquad
\mathbf B_x=\mathbf D^{-1}+\frac{\mathbf D^{-1}\mathbf g\mathbf g^{\mathrm{T}}\mathbf D^{-1}}{\kappa}.
\end{equation}
The second term is the positive semidefinite penalty on the known-clock covariance bound. Its rank is at most one for this axis. Thus,
\begin{equation}\label{eq:clockdiagonal}
B_{x,\ell\ell}=\frac1{d_\ell}+\frac{g_\ell^2}{d_\ell^2\kappa},\qquad
B_{x,\ell k}=\frac{g_\ell g_k}{d_\ell d_k\kappa}\ (\ell\ne k).
\end{equation}
For an existing path with fixed directional information, the derivative $\partial B_{x,\ell\ell}/\partial\kappa=-g_\ell^2/(d_\ell^2\kappa^2)\leq0$ shows the value of an additional common path. It raises the clock information available to an imbalanced coordinate, even if the new path never enters that coordinate's reciprocal product.
\begin{proposition}[Bilateral clock anchoring]\label{prop:bilateralanchor}
For either scalar clock offset, with path identities fixed in the isolated-path model, local identifiability holds if and only if at least one retained physical path has positive information in both directions. Equivalently, $\kappa>0$ in~\eqref{eq:clockschur}.
\end{proposition}
\begin{proof}[\hspace*{\parindent}Proof]
Each summand in~\eqref{eq:clockschur} is nonnegative and is positive exactly when both directional information quantities are positive. If all paths are one-sided, the Schur complement vanishes and a clock shift is absorbed by their physical coordinates as in~\eqref{eq:idmatrix}.
\end{proof}
For a one-sided path with $\kappa>0$,~\eqref{eq:clockdiagonal} reduces to $B_{x,\ell\ell}=\lambda_{q,\ell}^{-1}+\kappa^{-1}$. Local energy reduces the first term, while bilateral paths determine the clock variance floor. For a common path define $\kappa_\ell=4\lambda_{+,\ell}\lambda_{-,\ell}/d_\ell$, so $\kappa=\sum_j\kappa_j$. If independent, locally unbiased directional estimates attain their local bounds, $V_{\ell}^{\rm half}=\tfrac14(\lambda_{+,\ell}^{-1}+\lambda_{-,\ell}^{-1})=d_\ell^{-1}+g_\ell^2/(d_\ell^2\kappa_\ell)$. This follows from $d_\ell^2-g_\ell^2=4\lambda_{+,\ell}\lambda_{-,\ell}$, and comparison with~\eqref{eq:clockdiagonal} gives
\begin{equation}
V_{\ell}^{\rm half}-B_{x,\ell\ell}
=\frac{g_\ell^2}{d_\ell^2}
\left(\frac1{\kappa_\ell}-\frac1\kappa\right)\geq0.
\label{eq:halfsumjointgap}
\end{equation}
The gap vanishes for balanced information or a sole bilateral path. Another bilateral path makes it positive for an imbalanced coordinate. The gap quantifies the reduction in the variance bound from pooling clock information. The proposed receiver uses the common displacement in association and estimates physical parameters through pairwise coherent products.

For fixed information weights and $\kappa>0$, assume independent, unbiased local estimates with variances $\lambda_{q,\ell}^{-1}$ across all observed directions and paths. A common path yields the clock estimate $\widehat b_\ell=(\widehat\theta_{+,\ell}-\widehat\theta_{-,\ell})/2$ with variance $\kappa_\ell^{-1}$. Pooling with weights $\kappa_\ell/\kappa$ gives $\widehat b$ with variance $\kappa^{-1}$. For each path, $\widehat m_\ell=\sum_q\lambda_{q,\ell}\widehat\theta_{q,\ell}/d_\ell$ over available directions has variance $d_\ell^{-1}$. Its covariance with the local half difference vanishes for a common path because $\lambda_{q,\ell}\operatorname{var}(\widehat\theta_{q,\ell})=1$ in both directions. All half differences from other paths are independent of $\widehat m_\ell$. Hence $\widehat x_\ell=\widehat m_\ell-(g_\ell/d_\ell)\widehat b$ is unbiased and has the joint covariance in~\eqref{eq:clockbound}. The weights in this joint estimator are the directional information values of the isolated-path model. The clock contribution in~\eqref{eq:clockdiagonal} vanishes for a balanced path ($g_\ell=0$), and for any linear combination $\mathbf a^{\mathrm{T}}\widehat{\mathbf x}$ with $\mathbf a^{\mathrm{T}}\mathbf D^{-1}\mathbf g=0$. For two one-sided paths $i,j$, let $q_i,q_j\in\{+1,-1\}$ denote their receiving directions. Since $g_i/d_i=q_i$, the penalty on the variance bound for $x_i-x_j$ is $(q_i-q_j)^2/\kappa$. Two paths seen at the same station thus retain their known-clock precision for the difference, although each individual bound contains the clock variance. For paths seen at opposite stations the penalty is $4/\kappa$. For coupled paths and axes the corresponding condition is $\mathbf C^{\mathrm{T}}\mathbf F^{-1}\mathbf a=\mathbf0$ for the contrast vector $\mathbf a$.

\subsection{Directional Resources and Calibration}
For one isolated reciprocal path,~\eqref{eq:clockbound} reduces to
\begin{equation}
B_x=\frac14\left(\lambda_+^{-1}+\lambda_-^{-1}\right).
\label{eq:singlepairbound}
\end{equation}
This equals the physical half-sum variance when independent local estimates attain their directional bounds. At fixed total energy and equal apertures, write $\lambda_+=2\eta_E\lambda_0$ and $\lambda_-=2(1-\eta_E)\lambda_0$, where $\lambda_0$ is the balanced directional information. Relative to balance, the standard-deviation factor is
\begin{equation}
\sqrt{\frac{B_x(\eta_E)}{B_x(1/2)}}
=\frac{1}{2\sqrt{\eta_E(1-\eta_E)}}.
\label{eq:energyfactor}
\end{equation}
Directional power scales the local information through $\rho_{q,\ell}$. At fixed power per symbol,~\eqref{eq:lambda} and~\eqref{eq:uniformmoments} give $\lambda_\tau\propto M_q$ and $\lambda_\nu\propto M_q(M_q^2-1)$, so the Doppler bound is more sensitive to an imbalance in block lengths.

For unequal blocks, the measured-vector Doppler product retains $M_c=\min(M_A,M_B)$ samples and slow-time moment $S_{t,c}=T_s^2M_c(M_c^2-1)/12$, whereas scalar fusion uses both complete blocks. The fixed-identity raw-array bound includes cross-path coupling and uses the full aperture in each direction. Section~\ref{sec:numerical} varies directional energy and symbol allocation separately to examine these two effects.

In an idealized reference-aided control, an independent known reference observing the same clock adds information $\Lambda_r$ to $h$ in~\eqref{eq:kappa}. For a single one-sided target of information $\Lambda_i$, the resulting bound is $B_{x,i}^{\rm ref}=\Lambda_i^{-1}+\Lambda_r^{-1}$. This specialization explains the information supplied by the separately budgeted reference in Section~\ref{sec:numerical}.

\subsection{Residual Clocks Over Finite Blocks}
Constant opposite offsets cancel in the half-sum. Temporal evolution leaves a residual when reciprocal blocks sample different clock realizations. For $x\in\{\tau,\nu\}$, let $u_x,v_x$ be independent zero-mean, unit-variance Ornstein-Uhlenbeck (OU) processes with correlation time $T_c$ and $\rho_x(\Delta t)=e^{-|\Delta t|/T_c}$. Let $\sigma_x$ be the residual standard deviation and $\epsilon_x$ the common-component weight. The residual is
\begin{equation}\label{eq:oucomponents}
e_{x,q}(t)=\frac{\sigma_x[q\,u_x(t)+\epsilon_xv_x(t)]}{\sqrt{1+\epsilon_x^2}},
\quad q\in\{+1,-1\}.
\end{equation}
The cross-direction covariance is $\sigma_x^2\chi_x\rho_x(\Delta t)$ and $\chi_x=(\epsilon_x^2-1)/(\epsilon_x^2+1)$. TO enters as $-2\pi f_ne_{\tau,q}(t)$, whereas CFO accumulates into phase $\phi_q(t)=2\pi\int_0^t e_{\nu,q}(u)\,du$. Before compression, a reciprocal product contains the sum of these directional residual phases, so constant antisymmetric slopes cancel but evolving clocks generally do not.

For a fixed identity, let $\mathbf e_{x,q}=[e_{x,q}(t_{q,m})]_{m=0}^{M_q-1}$ be the sampled clock residuals and $\mathbf w_q$ the weights of a locally linear estimator. Write $\delta\widehat x=\tfrac12(\mathbf w_A^{\mathrm{T}}\mathbf e_{x,A}+\mathbf w_B^{\mathrm{T}}\mathbf e_{x,B})$, with $\mathbf1^{\mathrm{T}}\mathbf w_q=1$. Defining $[\mathbf K_{qr}]_{m,k}=e^{-|t_{q,m}-t_{r,k}|/T_c}$ gives
\begin{equation}
\begin{aligned}
\operatorname{var}(\delta\widehat x)=\frac{\sigma_x^2}{4}\bigl[
&\mathbf w_A^{\mathrm{T}}\mathbf K_{AA}\mathbf w_A
+\mathbf w_B^{\mathrm{T}}\mathbf K_{BB}\mathbf w_B\\
&+2\chi_x\mathbf w_A^{\mathrm{T}}\mathbf K_{AB}\mathbf w_B\bigr].
\end{aligned}
\label{eq:resblockvar}
\end{equation}
Uniform weights describe block-average delay perturbations. For a block-local left-integration benchmark on equally spaced samples, CFO produces phase $\phi_m=2\pi T_s\sum_{j=0}^{m-1}e_j+\phi_0$. Substitution into the least-squares phase slope and exchange of sums yield
\begin{equation}\label{eq:cfoweights}
w_j=\frac{T_s}{S_t}\sum_{m=j+1}^{M-1}(t_m-\bar t),
\quad j=0,\ldots,M-2,
\end{equation}
where $S_t=\sum_m(t_m-\bar t)^2$ and $w_{M-1}=0$. Centered times sum to zero, removing $\phi_0$ and all pre-block phase. These weights sum to one and hence reproduce a constant CFO. For a varying CFO they give the response of the ordinary least-squares slope to each frequency sample. Substituting them into~\eqref{eq:resblockvar} gives a finite-block reference for the nonlinear receiver.

Selecting one sample in each direction and taking their half-sum $\bar e_x$ gives
\begin{equation}
\operatorname{var}(\bar e_x)=\frac{\sigma_x^2}{2}
[1+\chi_x\rho_x(\Delta t)].
\label{eq:resvar}
\end{equation}
At equal times the antisymmetric component cancels. Adjacent blocks have $\Delta t=M_AT_s+T_{\rm gap}$ between equal-index samples for interblock gap $T_{\rm gap}$, leaving a residual even at $\epsilon_x=0$. Equation~\eqref{eq:resblockvar} gives the variance for the stated linear response. Section~\ref{sec:numerical} tests the complete receiver.

\section{Numerical Results and Discussion}\label{sec:numerical}
We examine how selecting the common paths affects the accuracy of reciprocal estimates, then vary the energy, observation duration, and clock stability of the two links.

\subsection{Simulation Setup and Evaluation Metrics}
Table~\ref{tab:simparams} gives the nominal waveform and array settings. For baseline length $d_{AB}=0.60cT_{\rm CP}$, targets are drawn uniformly over $x\in[0.1d_{AB},0.9d_{AB}]$ and $|y|\in[0.0255,0.4675]cT_{\rm CP}$ with equiprobable signs. Selected paths are geometrically separated and differ in bistatic range by at least $1.5\delta_r$, where $\delta_r=c/(2N\Delta f)$, with $R<cT_{\rm CP}-2\delta_r$. A target is stationary with probability $0.3$. Otherwise its speed and heading are uniform on $[10,25]$ m/s and $[0,2\pi)$. Received power scales as $R^{-2.2}$. At nominal allocation a common path has equal power and independent phases in the two directions. SNR is the summed common-path power at one antenna divided by the noise variance. Private paths add received energy. All four visibility patterns have three common paths. S1 has no private paths, S2 adds one per station, and S3 and S4 add two at BS A or BS B, respectively.

Each receiver knows its local path count. The angle search selects that many peaks using subspace fit and peak strength. Association uses the gates in Table~\ref{tab:simparams} and admits delay and Doppler half-sums in $[-0.02T_{\rm CP},1.02T_{\rm CP}]$ and $[-1.10\nu_{\max},1.10\nu_{\max}]$, where $\nu_{\max}=2f_c(40\,\mathrm{m/s})/c$. The geometry term is used for well-conditioned intersections.

Root mean square error (RMSE) is computed over finite outputs against the corresponding BS A targets, so a wrong BS B partner still contributes to the error. Precision is the fraction of returned pairs with the correct common identity. Recall is the fraction of common paths correctly paired. Offline scoring assigns local identities by one-to-one direction-cosine matching with gate $0.04$. All recovery rates use planned targets as the denominator. $P_{\rm out}$ requires finite outputs on both axes, $P_{\rm tol}$ also requires both physical tolerances in Table~\ref{tab:simparams}, and $P_{\rm use}$ further requires the correct reciprocal identity. Methods reporting estimates solely at BS A are evaluated with the first two rates. The accuracy and clock tests score the two longer paths. The association, ablation, and resource tests score all three. All paths remain in the observations and the Fisher information calculation.
\begin{table}[t]
\caption{Simulation Parameters}
\label{tab:simparams}
\centering
\footnotesize
\begin{tabular}{@{}ll@{}}
\toprule
Parameter & Value\\\midrule
Carrier frequency; receive elements & 26 GHz; 16\\
Active subcarriers; spacing & 180; 60 kHz\\
FFT size; sampling rate & 540; 32.4 MHz\\
CP length; symbol interval & 39 samples; $17.87037\,\mu$s\\
Nominal symbols, BS A / BS B & 280 / 280\\
Node separation; noise variance & 216.517 m; 1\\
Constant TO/CFO standard deviations & 20 ns / 100 Hz\\
Range/Doppler error tolerances & 0.5 m / 1 Hz\\
Common paths; local orders & 3; $(3,3),(4,4),(5,3),(3,5)$\\
Consensus widths; rejection cost & 8 ns / 25 Hz; $\gamma=1$\\
Cost $(w_p,\lambda_r,\lambda_Q)$ & $(0.05,0.30,0.05)$\\
Score floor $q_0$; power scale floor & 0.1; 0.1\\
Penalties $\lambda_p,\lambda_b,\lambda_g$ & $10^5$ each\\
Geometry $(w_g,s_g,g_{\max})$ & $(1,10\,\mathrm m,2\delta_r)$\\
LCMV $(\epsilon_{\rm rel},\epsilon_{\rm abs},\epsilon_G)$ & $(0.01,10^{-12},10^{-12})$\\
Weak-reference threshold $\epsilon_{\rm ref}$ & $10^{-8}$\\
Iterations: AIQML / consensus / search & 3 / 5 / 10\\
\bottomrule
\end{tabular}
\end{table}
\begin{figure*}[!t]\centering\includegraphics[width=0.92\textwidth]{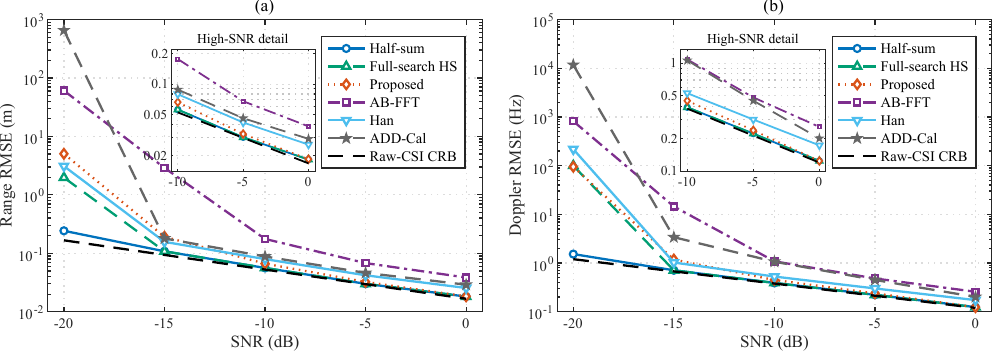}\caption{\CapBasic}\label{fig:basic}\end{figure*}

The accuracy and association sweeps span $-20$ to $0$ dB in 5 dB steps. The association curves use 50 scenes per SNR and visibility pattern. We repeat the comparisons on 50 independently generated scenes at $-20$, $-15$, and $-10$ dB, using the same scenes across the four visibility patterns and keeping receiver settings fixed. Pointwise 95\% intervals use 4000 paired resamples of scenes, keeping all compared conditions from a scene together. Ratios on common reported targets use the intersection of the reported sets.
\begin{figure*}[!t]
\begin{minipage}[t]{\columnwidth}
\centering
\includegraphics[trim=0 10pt 0 0,clip,width=\linewidth]{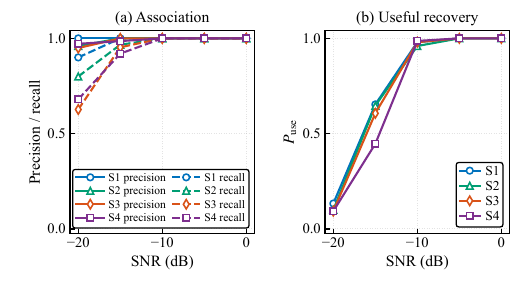}
\vspace{-11pt}
\caption{\CapAssociation}\label{fig:association}
\end{minipage}\hfill
\begin{minipage}[t]{\columnwidth}
\centering
\includegraphics[trim=0 10pt 0 0,clip,width=\linewidth]{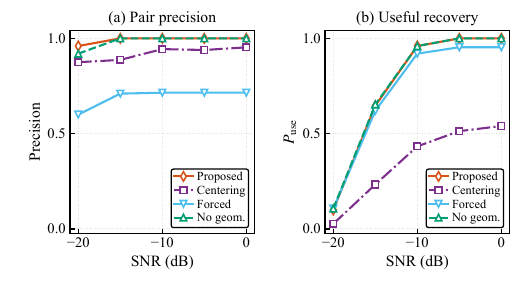}
\vspace{-11pt}
\caption{\CapAblation}\label{fig:ablation}
\end{minipage}
\vspace{-10pt}
\end{figure*}
\subsection{Estimation and Association Performance}
In Fig.~\ref{fig:basic}, \emph{Proposed} uses PreVote for association and the paired phase vectors for recovery. Han applies bistatic signal matching and matrix pencil estimation~\cite{han2025ota} to target echoes from both directions. Our implementation of Han pairs responses by delay rank and pools their TO and CFO estimates by an energy-weighted median. ADD-Cal adapts the delay and phase correction in~\cite{ding2025bistatic} with 280 additional reference symbols per direction on a static path of known delay. After calibration, both methods use MUSIC, LCMV, and AIQML to estimate range and Doppler from target echoes at BS A. The Raw-CSI CRB is the square root of the mean finite bound in~\eqref{eq:schur} with true identities and fixed phase branches.

At $-20$ dB the proposed receiver, Han, and ADD-Cal report 345, 391, and 308 of 400 targets, respectively. Han gives the lowest range and Doppler RMSE on the 276 targets reported by all three. At $-15$ dB the proposed receiver approaches Han's accuracy on the 398 common outputs, with RMSEs of $0.192$ m and $1.180$ Hz versus $0.156$ m and $1.023$ Hz for Han. At $-10$, $-5$, and $0$ dB all three report all 400 targets, and the proposed receiver has the lowest RMSE on both axes. This higher-SNR advantage is consistent with how the receivers handle clock errors. In a calibrated observation at BS A, any residual timing or frequency offset enters the fitted delay or Doppler through~\eqref{eq:effective}. For a common path, the reciprocal product in~\eqref{eq:product} cancels the opposite constant clock terms before the physical slope is fitted. The fit therefore uses both directional phase measurements without inheriting error from a separate clock correction. At $-10$ dB reciprocal fusion yields $0.0665$ m and $0.4494$ Hz, compared with $0.0796$ m and $0.5283$ Hz for Han and $0.0883$ m and $1.0653$ Hz for ADD-Cal. At $0$ dB both RMSEs are about $28\%$ lower than Han's.\par
AB-FFT fits the same paired products with a windowed FFT and phase refinement ($0.1762$ m and $1.0865$ Hz at $-10$ dB). Half-sum averages the accepted local slopes by~\eqref{eq:scalarfusion}, while Full-search HS searches all displacements before averaging. At $0$ dB the paired 95\% intervals for the proposed-to-Half-sum RMSE ratios are $[0.990,1.025]$ and $[0.983,1.015]$ for range and Doppler. The proposed receiver comes within $10\%$ of the Raw-CSI CRB on both axes with $0.172\%$ of the raw S1 payload. The paired phase vectors thus give accuracy close to the local slope average at this operating point.\par
Fig.~\ref{fig:association} shows that pair precision and recall reach one at $-10$ dB in all four visibility patterns. In a separate comparison on independently generated scenes, PreVote and Full-search each identify $597$ of $600$ common paths without a false pair, and 573 meet both physical tolerances for each method. Across 600 combinations of scene, SNR, and visibility, PreVote reduces median association time from $5.229$ to $0.966$ ms and total receiver-stage time from $132.29$ to $130.65$ ms. At $-20$ dB the two rules correctly identify $429$ and $392$ of $441$ accepted pairs each, a precision gain of $8.39$ percentage points (95\% interval $[5.36,11.93]$). Of 600 planned targets, 59 and 55 also meet both error tolerances. In S2 the sufficient conditions following~\eqref{eq:assign} hold in 19, 46, and 50 of the 50 independent scenes at $-20$, $-15$, and $-10$ dB, respectively. All three true pairs survive in every such scene and in one additional scene at $-20$ dB. Selecting the displacement before assignment thus preserves the recovery count of Full-search at $-10$ dB with less time spent on association.

Fig.~\ref{fig:ablation} compares association rules in S2 using the same local features and coherent recovery stage. At $-15$ dB, PreVote correctly pairs and recovers 97 of 150 planned common targets within both tolerances, compared with 35 for Centering. On 50 independently drawn scenes the counts are 95 and 40 out of 150. The respective gains are 41.3 and 36.7 percentage points (95\% intervals $[31.3,51.3]$ and $[27.3,46.0]$). PreVote estimates displacement from candidate pairs, while Centering uses the list medians, which can differ across visible path sets by~\eqref{eq:centeringbias}. On the plotted scenes, forcing four pairs caps S2 precision at $3/4$ (observed $0.7150$ at $0$ dB). Removing the geometry check at $-20$ dB lowers precision from $0.9600$ to $0.9197$ and adds one useful recovery. On the independent scenes at $-15$ dB, all 145 pairs accepted by PreVote are correct, compared with 55 of 64 for Centering. Both rules give identical estimates for their 54 shared correct pairs. The gain in useful recovery therefore comes from identifying more common paths whose paired measurements meet both error tolerances.

\makeatletter
\newcommand{\ResourceSubsection}{\@startsection{subsection}{2}{\z@}%
  {2ex minus 0.5ex}{0.7ex plus 0.5ex}{\normalfont\normalsize\itshape}}
\makeatother
\ResourceSubsection{Directional Resources and Clock Robustness}
\begin{figure*}[t]\centering\includegraphics[width=0.92\textwidth]{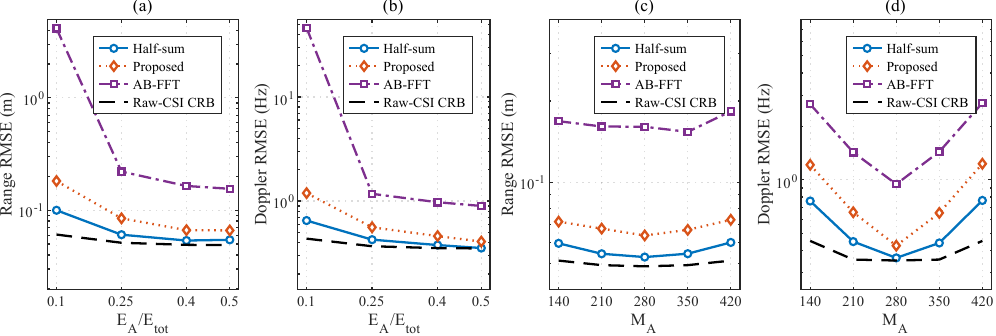}\caption{\CapResources}\label{fig:resources}\end{figure*}
\begin{figure}[t]\centering\includegraphics[width=\columnwidth]{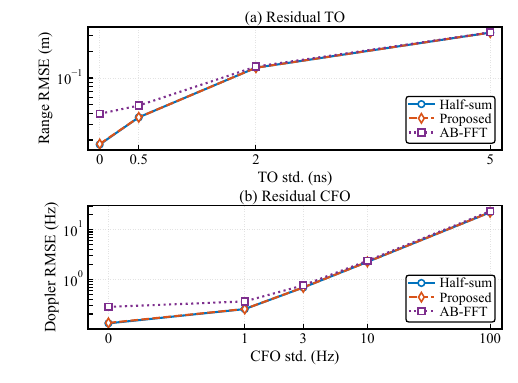}\vspace{-8pt}\caption{\CapResidual}\label{fig:residual}\vspace{-15pt}\end{figure}
In S1, Fig.~\ref{fig:resources} varies directional resources at $-10$ dB over 400 scenes. The energy sweep holds $M_A=M_B=280$ and total received energy fixed, using directional powers $2\eta_E\rho$ and $2(1-\eta_E)\rho$ for $\eta_E\in\{0.1,0.25,0.4,0.5\}$. Balancing the links ($\eta_E=0.1$ to $0.5$) reduces range RMSE from $0.1819$ to $0.0661$ m and Doppler RMSE from $1.1933$ to $0.4088$ Hz. Useful recovery rises from $0.6125$ to $0.9825$. A weak direction increases the clock penalty in~\eqref{eq:clockdiagonal} for an imbalanced path and the relative product error in~\eqref{eq:normalizednoise}. For Half-sum, the ratio of Doppler RMSE at $\eta_E=0.1$ to that at balance is $1.837$ $[1.735,1.944]$ on common reported targets. The isolated-pair benchmark gives $1.667$ by~\eqref{eq:energyfactor}. At fixed total energy, strengthening the weaker direction reduces the influence of the noisier input on the reciprocal product.

The symbol sweep fixes power per symbol and $M_A+M_B=560$, with $M_A\in\{140,210,280,350,420\}$. Changing the symbol allocation from $140/420$ to $280/280$ reduces Doppler RMSE for the proposed receiver from $1.2147$ to $0.4258$ Hz and raises $P_{\rm use}$ from $0.6142$ to $0.9742$. The raw Doppler information scales as $M_q(M_q^2-1)$ by~\eqref{eq:uniformmoments}. Each range vector uses all local symbols, but reciprocal Doppler recovery uses only the first $M_c$ samples from each block. At $140/420$ the ratio of the proposed receiver's Doppler RMSE to that of Half-sum is $1.599$ $[1.523,1.680]$, compared with approximately $1.20$ for range. At fixed raw-CSI volume, increasing $M_c$ from 140 to 280 raises the S1 vector payload $B_{\rm vec}$ from 30\,720 to 44\,160 bytes and leaves the association message at 240 bytes. Reassigning symbols from the longer block to the shorter one extends their common aperture up to balance, explaining the measured improvement in reciprocal Doppler accuracy.

Fig.~\ref{fig:residual} evaluates clock variation at $0$ dB over 200 scenes ($T_c=0.05$ s, $\epsilon_x=0.1$). With 280 symbols per block and a $0.5$ ms gap, equal indices are $5.504$ ms apart. CFO phase is integrated from the left on the combined grid. The proposed receiver attains $P_{\rm use}=1$ at $0.5$ and $2$ ns TO and at $1$ Hz CFO. At $5$ ns TO, range RMSE reaches $0.3256$ m ($0.9994$ times the Half-sum RMSE, interval $[0.9984,1.0004]$). All targets are correctly paired and reported, so the decline to $P_{\rm use}=0.870$ comes entirely from physical estimation error. At $10$ and $100$ Hz CFO, Doppler RMSE reaches $2.2536$ and $22.514$ Hz. The finite-block prediction of $0.2194\sigma_\nu$ in~\eqref{eq:resblockvar}--\eqref{eq:cfoweights} gives $2.194$ and $21.942$ Hz, closer at both levels than the instantaneous prediction of $0.247\sigma_\nu$ in~\eqref{eq:resvar}. Phase accumulated before a block affects its intercept, whereas frequency fluctuations within it affect the fitted slope. Accounting for that distinction predicts the scale of the residual Doppler error at both larger CFO levels.

\section{Conclusion}\label{sec:conclusion}
\begingroup\widowpenalty=10000
In asynchronous bistatic OFDM sensing, physical range and Doppler can be recovered from reciprocal observations of common paths. To find those paths under partial visibility, the proposed receiver selects pairs that support a common clock displacement, then combines their coherently compressed phase samples. The opposite constant clock terms cancel in each reciprocal product. For a correctly associated pair, the remaining signal slopes are twice the delay and Doppler phase slopes of one path. With path identities and phase branches fixed, the joint CRB shows how common paths constrain the clock and how its uncertainty couples the physical estimates across paths. Balancing the energy between directions improved recovery at fixed total energy, and extending the common slow-time aperture improved Doppler accuracy at fixed total symbol count. With a varying carrier-frequency offset, the residual Doppler error reflected the phase evolution within the two observation blocks.

\par\endgroup

\begingroup
\renewcommand{\IEEEbibitemsep}{4.235pt plus 0.18pt minus 0.12pt}
\providecommand{\BIBdecl}{\interlinepenalty=10000\relax}
\IEEEtriggercmd{\newpage\setlength{\itemsep}{2.494pt plus 0.18pt minus 0.12pt}}
\IEEEtriggeratref{17}
\bibliographystyle{IEEEtran}
\bibliography{references}
\endgroup
\end{document}

%% file: expanded_matrix.tex
\begin{figure*}[!b]
\normalsize
\renewcommand{\arraystretch}{1.05}
\setlength{\arraycolsep}{4pt}
\centering
\noindent\rule{\textwidth}{0.4pt}\vspace{-2pt}
\setcounter{ExpandedMatrixSavedEquation}{\value{equation}}
\setcounter{equation}{\numexpr\ExpandedMatrixNumber-1\relax}
\renewcommand{\theHequation}{expandedmatrix}
\begin{equation}
\begin{aligned}
\mathbf X_{q,i_q}
&=\beta_{q,i_q}
\begin{bmatrix}
1 & e^{j2\pi T_s(\nu_\ell+qf_o)} & \cdots & e^{j2\pi(M-1)T_s(\nu_\ell+qf_o)}\\
e^{-j2\pi\Delta f(\tau_\ell+q\tau_o)} & e^{j2\pi[T_s(\nu_\ell+qf_o)-\Delta f(\tau_\ell+q\tau_o)]} & \cdots & e^{j2\pi[(M-1)T_s(\nu_\ell+qf_o)-\Delta f(\tau_\ell+q\tau_o)]}\\
\vdots & \vdots & \ddots & \vdots\\
e^{-j2\pi(N-1)\Delta f(\tau_\ell+q\tau_o)} & e^{j2\pi[T_s(\nu_\ell+qf_o)-(N-1)\Delta f(\tau_\ell+q\tau_o)]} & \cdots & e^{j2\pi[(M-1)T_s(\nu_\ell+qf_o)-(N-1)\Delta f(\tau_\ell+q\tau_o)]}
\end{bmatrix}
+\mathbf E_{q,i_q}.
\end{aligned}
\label{eq:expandedmatrix}
\end{equation}
\vspace{-3pt}
\setcounter{equation}{\value{ExpandedMatrixSavedEquation}}
\end{figure*}